\documentclass[]{aa}
\usepackage{hyperref}
\usepackage{graphicx}
\usepackage{multirow}
\usepackage{txfonts}
\usepackage{lipsum}

\bibpunct{(}{)}{;}{a}{}{,}

\newcommand{\paperone}{\mbox{Paper I}}
\newcommand{\papertwo}{\mbox{Paper II}}
\newcommand{\lya}{Ly$\alpha$}
\newcommand{\lyb}{Ly$\beta$}
\newcommand{\rsun}{$R_\odot$}
\newcommand{\avg}[1]{\langle#1\rangle}

\begin{document}

\title{Hot prominence detected in the core of a Coronal Mass Ejection: \\ III. Plasma filling factor from UVCS Lyman-$\alpha$ and Lyman-$\beta$ observations}
\titlerunning{TBD}
\author{R. Susino\inst{1} \and A. Bemporad\inst{1} \and S. Jej\v ci\v c\inst{2,3} \and P. Heinzel\inst{3}}
\authorrunning{R. Susino et al. }
\institute{INAF -- Turin Astrophysical Observatory, 10025 Pino Torinese (TO), Italy \and Faculty of Mathematics and Physics, University of Ljubljana, 1000 Ljubljana, Slovenia \and Astronomical Institute, The Czech Academy of Sciences, 25165 Ond\v rejov, Czech Republic}
\abstract{This work deals with the study of an erupting prominence embedded in the core of a coronal mass ejection that occurred on August 2, 2000, and focuses on the derivation of the prominence plasma filling factor.}{We explore two methods to measure the prominence plasma filling factor along the line of sight that are based on the combination of visible-light and ultraviolet spectroscopic observations.}{Theoretical relationships for resonant scattering and collisional excitation are used to evaluate the intensity of the neutral hydrogen Lyman-$\alpha$ and Lyman-$\beta$ lines, in two prominence points where simultaneous and cospatial LASCO-C2 and UVCS data were available. Thermodynamic and geometrical parameters assumed for the calculation (i.e., electron column density, kinetic temperature, flow velocity, chromospheric \lya\ and \lyb\ intensities and profiles, and thickness of the prominence along the line of sight) are provided by both observations and the results of a detailed 1D non-LTE radiative-transfer model of the prominence, developed in our previous work \citep{heinzel2016}. The geometrical filling factor is derived from the comparison between the calculated and the measured intensities of the two lines. The results are then checked against the non-LTE model in order to verify the reliability of the described methods.}{The resulting filling factors are consistent with the model in both the prominence points when the separation of the radiative and collisional components of the total intensity of the hydrogen lines, required to estimate the filling factor, is performed using the both the \lya\ and \lyb\ line intensities. The exploration of the parameter space shows that the results are weakly sensitive to the plasma flow velocity, while they depends more strongly on the assumed kinetic temperatures.}{The combination of visible-light and ultraviolet \lya\ and \lyb\ data can be used to approximately estimate the line-of-sight geometrical filling factor in erupting prominences, but the proposed technique, which is model-dependent, is reliable only for emission that is optically thin in the lines considered, condition that is not in general representative of prominence plasma.}

\maketitle

\section{Introduction}\label{intro}
Prominences are underlying structures of the solar atmosphere consisting of cool and dense plasma (typically at temperatures of~\mbox{$\sim 10^4$~K} and densities of~$\sim 10^{11}$~cm$^{-3}$) sustained against the solar gravity by the magnetic field---they are thought to be made of chromospheric gas entrapped in magnetic flux ropes anchored on the Sun's surface \citep[see][for a comprehensive treatment on solar prominences]{mackay2010,labrosse2010,vial2015}.
They can eventually erupt due to magnetic instabilities that trigger coronal mass ejections \citep[CMEs; see][]{chen2011,webb2012}, appearing then as the bright core of these structures in visible-light images acquired by coronagraphs \citep[e.g.,][]{akmal2001,ciaravella2003}.

Observations, carried out in particular in the ultraviolet wavelength domain \citep[see, e.g.,][for a review on SOHO observations of prominences]{patsourakos2002}, have provided evidence that they are made up of small-scale, elongated threads and fine structures filling only a small fraction of the prominence volume \citep[e.g.,][]{berger2014}.
Thermodynamic modeling of prominences, which is based on the measured emission, critically depends on the proper knowledge of the real volume filled by the radiating plasma \citep{labrosse2010,labrosse2015}.
The ``filling factor'' is the crucial parameter that gives a measure of the effective emitting volume and, therefore, it is fundamental for a correct interpretation of observed line emission.

There are several definitions of filling factor and many ways to measure it \citep[see the discussion in][]{labrosse2010}.
For the optically-thin emission coming from the transition-region envelop of prominences \citep[the so-called PCTR; see][]{parenti2015}, the most common approach is to divide the inferred emission measure by the square of the electron density, derived from the ratio of density-sensitive lines, and compare the result with the prominence size estimated from observations \citep[see, e.g.,][]{mariska1979}. 
Resulting filling factors are of the order of a few percents \citep[up to $\sim 0.03$; see, e.g.,][]{labrosse2010} and suggest highly inhomogeneous density distributions.
The prominence cool counterpart can be more structured than the PCTR, implying values of the filling factor even lower.

An alternative method that can be used to estimate this parameter is based on the ratio of the intensity of a collisionally excited emission line to the square of the visible-light polarized brightness ($pB$).
This approach was used by \citet{fineschi1994} and \citet{romoli1994} to derive the coronal irregularity factor \citep[which is actually proportional to the reciprocal of the filling factor; see][]{allen1963} from the relative intensities of the neutral hydrogen Lyman-$\alpha$ and Lyman-$\beta$ emission lines, for a number of coronal structures.

In this work we explore a similar technique, showing that an estimate of the filling factor can be obtained from simultaneous and cospatial visible-light and ultraviolet observations of an erupting prominence embedded in the core of a CME.
This event was already studied in a couple of papers \citep[][hereafter \paperone\ and \papertwo, respectively]{heinzel2016,jejcic2017}, where SOHO/LASCO-C2 and SOHO/UVCS data were used to constrain a thermodynamic model of the prominence structure.
In comparison with standard quiescent prominences, this one turned out to be relatively hot and tenuous (with temperature~$\sim 10^5$~K and electron density~$\sim 10^8$~cm$^{-3}$) because of its expansion at quite large velocities.
In the present analysis, we exploit UVCS and LASCO-C2 visible-light observations, as well as the modeling results, to test this alternative way to infer the prominence plasma filling factor.

The paper is organized as follows: in Section~\ref{theory} we recall the theoretical formulation of resonant scattering and collisional excitation of coronal neutral hydrogen Lyman lines; in Section~\ref{observations} we summarize the observational and modeling results obtained in \paperone; we then describe the analysis methods and results in Sections~\ref{analysis} and~\ref{results}, and discuss them in Section~\ref{discussion}.

\section{Basic theory of UV line formation}\label{theory}
Under typical coronal conditions, the upper levels of the neutral hydrogen \lya\ and \lyb\ transitions are populated mainly through the mechanisms of photon absorption and collisional excitation and depopulated by spontaneous emission toward the ground level.
Therefore, the total integrated intensity in the lines is a mixture of radiative (i.e., produced by resonant scattering) and collisional components:
\begin{equation}\label{eq:i_tot}
I_\text{tot}=I_\text{rad}+I_\text{col}.
\end{equation}

According to \citet{noci1987}, the radiative component is given by
\begin{equation}\label{eq:i_rad}
I_\text{rad}=B_{ij}\,h\lambda_{ij}\frac{b_{ij}}{4\pi}\int_\text{LOS}n_i\int_\Omega p(\phi)F(\delta\lambda)\,d\omega\,dl,
\end{equation}
where $B_{ij}$ is the Einstein coefficient for photon absorption from the ground level $i$ to the excited level $j$ (in units of sr cm$^2$ erg$^{-1}$ s$^{-1}$), $h$ the Planck constant, $\lambda_{ij}$ the rest wavelength of the transition, $b_{ij}$ the branching fraction for de-excitation ($b_{12}=1$ for Ly$\alpha$ and $b_{13}\simeq0.88$ for Ly$\beta$), $n_i$ the number density of hydrogen atoms in the ground level, $\Omega$ the solid angle subtended by the incident radiation source, $p(\phi)$ gives the angular dependence of the scattering process, and
\begin{equation}\label{eq:dimming_factor}
F(\delta\lambda)=\int_0^\infty I_\odot(\lambda-\delta\lambda)\Phi(\lambda)\,d\lambda,
\end{equation}
is the so-called Doppler-dimming factor, which accounts for the velocity-dependent irradiation and depends on the Doppler shift of the incident radiation as seen by the scattering atom moving with velocity $v$ along the direction of illumination, $\delta\lambda=\lambda_{ij}v/c$, the spectral profile of the incident radiation, $I_\odot(\lambda)$ (in units of erg cm$^{-2}$ s$^{-1}$ sr$^{-1}$ \AA$^{-1}$), and the normalized absorption profile, $\Phi(\lambda)$. 
If the velocity distribution of the scattering atoms is Maxwellian, the absorption profile is Gaussian with Doppler width given by 
\begin{equation}\label{eq:d_lambda}
\Delta\lambda_{D}=\frac{\lambda_{ij}}{c}\sqrt{\frac{2k_BT}{m}},
\end{equation}
where $k_B$ is the Boltzmann constant, $T$ is the kinetic (or ion) temperature, and $m$ is the atomic mass.
However, non-thermal motions, such as bulk expansion and turbulence, can broaden the absorption profile and increase the effective observed plasma temperature so that $T_\text{eff}=T+m\,\xi^2/(2k_B)$, being $\xi$ the Gaussian-distributed non-thermal velocity.

For a low-density coronal plasma, the number density of hydrogen atoms in the ground level can be approximated in the following way:
\begin{equation}\label{eq:ioniz}
n_i\approx n_H\approx 0.83\,R(T)\,n_e,
\end{equation} 
where $n_H$ is the number density of neutral hydrogen, the factor 0.83 is the ratio between proton and electron density for a fully ionized gas with 10\% helium, and $R(T)$ is the hydrogen ionization fraction at temperature $T$.

In Equation~(\ref{eq:i_rad}), integration is performed along the line of sight (LOS) across all the corona.
However, when the emitting plasma is confined in a small region where all the thermodynamic quantities can be reasonably considered uniform and approximated by their mean values, and the contribution from the surrounding corona can be neglected, using Eq.~(\ref{eq:ioniz}) the radiative component can be simplified as
\begin{equation}\label{eq:i_rad_approx}
I_\text{rad}\approx B_{ij}\,h\lambda_{ij}\frac{b_{ij}}{4\pi}\,\cdot\frac{\avg{\Omega}}{4\pi}F(\avg{w})\cdot0.83\,R(\avg{T})\avg{n_e}\cdot D_\text{eff},
\end{equation}
where $D_\text{eff}$ is the effective length of the LOS section across the emitting plasma volume and the brackets denote LOS-averaged values.
The factor $\avg{\Omega}/4\pi$ is the dilution factor.

The collisional component has the following form:
\begin{equation}\label{eq:i_col}
I_\text{col}=\frac{hc}{\lambda_{ij}}\frac{b_{ij}}{4\pi}\int_\text{LOS}n_i\,q_{ij}(T)\,n_e\,dl,
\end{equation}
where $q_{ij}(T)\cdot n_e$ is the collisional emission rate coefficient at temperature $T$.
With the same considerations as above, the previous expression can be approximated as
\begin{equation}\label{eq:i_col_approx}
I_\text{col}\approx\frac{hc}{\lambda_{ij}}\frac{b_{ij}}{4\pi}\cdot q_{ij}(\avg{T})\cdot0.83\,R(\avg{T})\avg{n_e^2}\cdot D_\text{eff}.
\end{equation}

Introducing the electron column density \mbox{$N_e\approx\avg{n_e}\cdot D_\text{eff}$,} giving the total number of electrons per unit area along the line of sight, and the irregularity factor \mbox{$X\equiv\avg{n_e^2}/\avg{n_e}^2$}, giving a measure of the inhomogeneity of the coronal electron density distribution \citep{fineschi1994}, it follows that \mbox{$I_\text{rad}\propto N_e$} and \mbox{$I_\text{col}\propto N_e^2\cdot X/D_\text{eff}$}. 
Since the effective thickness of the emitting plasma along the LOS and the irregularity factor are usually unknown, all information about the LOS plasma distribution can be conveniently enclosed in the so-called filling factor $f$, which is defined so that
\begin{equation}\label{eq:ff}
D_\text{eff}=D\cdot X \cdot f,
\end{equation}
where the length $D$ is the apparent thickness of the emitting feature.
In this way, it is possible to account for inhomogeneities of the coronal electron-density distribution and potential fragmentations of the plasma along the LOS, since $f$ gives a measure of the real fraction of the LOS filled with the plasma emitting the observed spectral lines, in case that $D$ is over/underestimated.
Note that usually filling factor calculations assume that the emitting plasma consists of material of uniform density surrounded by empty space, which is a simplification implying $X\approx 1$ \citep[see][]{labrosse2010}.
Note also that, of the two intensity components mentioned above, only the collisional one explicitly depends on the filling factor, through its dependence on effective thickness and irregularity factor. 

Combination of Equations~(\ref{eq:i_rad_approx}),~(\ref{eq:i_col_approx}), and~(\ref{eq:ff}) can be used to derive the filling factor from the observed intensities of the hydrogen \lya\ and \lyb\ lines if the electron column density is measured, for instance, from visible-light data:
\begin{equation}\label{eq:ff_calc}
f=K(T)\cdot\frac{N_e^2}{I_\text{col}\cdot D}=K(T)\cdot\frac{N_e^2}{(I_\text{tot}-I_\text{rad})\cdot D},
\end{equation}
where
\begin{equation}
K(T)=\frac{hc}{\lambda_{ij}}\frac{b_{ij}}{4\pi}\cdot q_{ij}(\avg{T})\cdot0.83\,R(\avg{T})
\end{equation} 
depends on the average kinetic temperature and the other atomic parameters.

We finally emphasize that with the approximations used in the above relationships all the measured and derived quantities must be regarded as LOS weighted averages; moreover, each term in the equations (Doppler-dimming coefficient, collisional rates, hydrogen ionization fraction, etc.) has a different weighting with electron density, temperature, and flow velocity, thus they may be representative of different parts of the emitting structure if the distribution of the plasma parameters along the LOS is not actually uniform.

\section{Prominence observations and modeling results}\label{observations}
\begin{figure}
\includegraphics[width=\columnwidth]{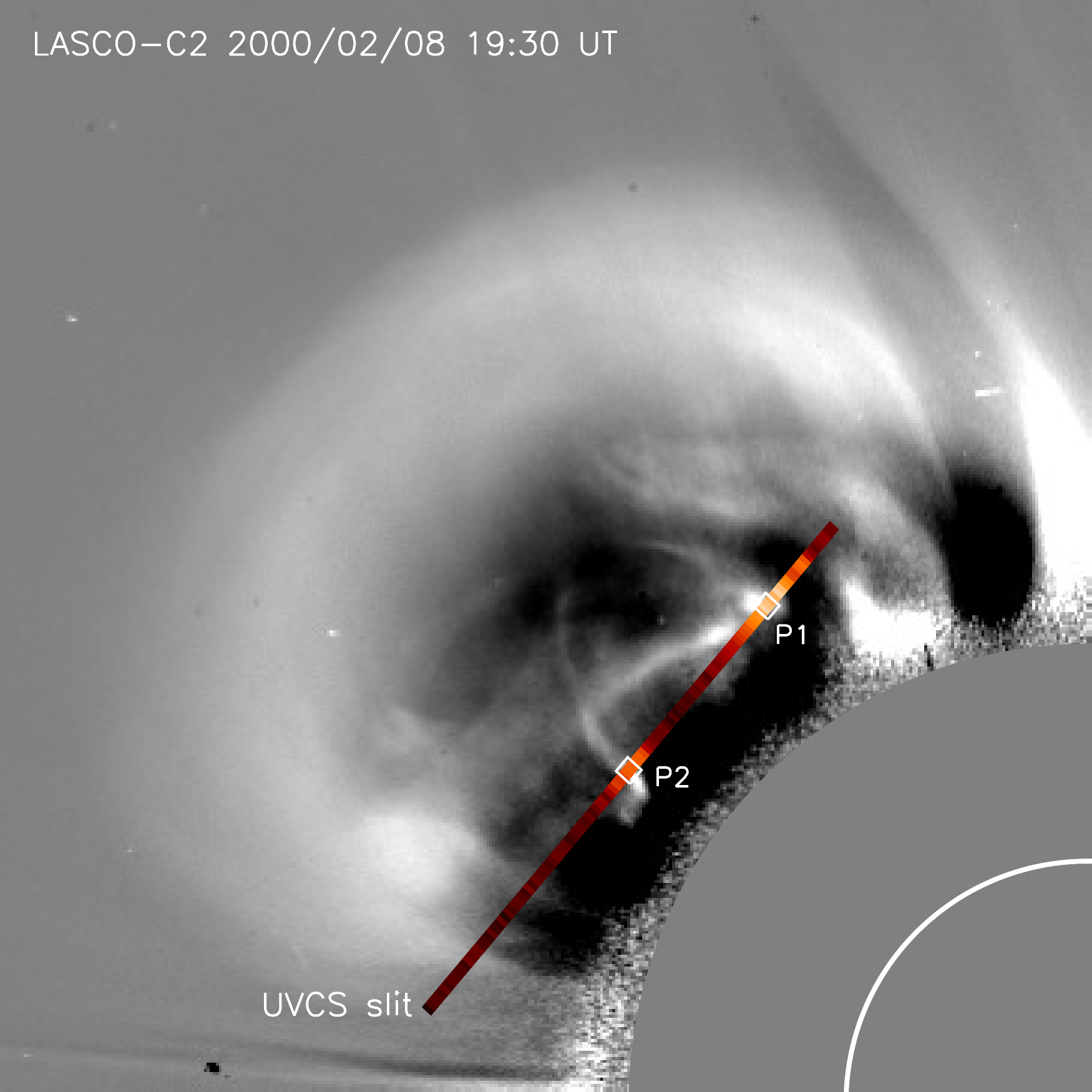}
\caption{\label{fig:lasco}LASCO-C2 mass image acquired at 19:30~UT on August 2, 2000, and intensity distribution of the \lya\ line along the UVCS slit (measured at the same time and represented with color gradient). The two points considered in this analysis (P1, located at a helio-latitude of~$\sim57^\circ$N, and P2, located at $\sim37^\circ$N) are also indicated.}
\end{figure}

In \paperone\ we presented observations of an erupting prominence in the core of a CME occurred on 2000 August 2. We refer the reader to \paperone\ and \papertwo\ for a detailed description of the event.
The prominence was observed in the visible light by the LASCO-C2 coronagraph and in the UV by the UVCS spectro-coronagraph, both on board the SOHO spacecraft.

Total-brightness LASCO images were used to infer the electron column density $N_e$ of the CME and prominence plasma by applying the method described in \citet{vourlidas2000} to the so-called ``excess-brightness'' image of the CME (see Figure~\ref{fig:lasco}), obtained by subtracting from the LASCO-C2 frame containing the CME at 19:30~UT, the pre-event frame acquired at 16:04~UT.
LASCO images were also used to measure the plane-of-the-sky (POS) velocity of the prominence, $v_\text{POS}\simeq300$~km~s$^{-1}$, as well as its projected thickness, $D\simeq56000$~km, which was taken as an approximation of the prominence average LOS thickness, under the simple assumption of cylindrical geometry.

\begin{figure}
\includegraphics[width=\columnwidth]{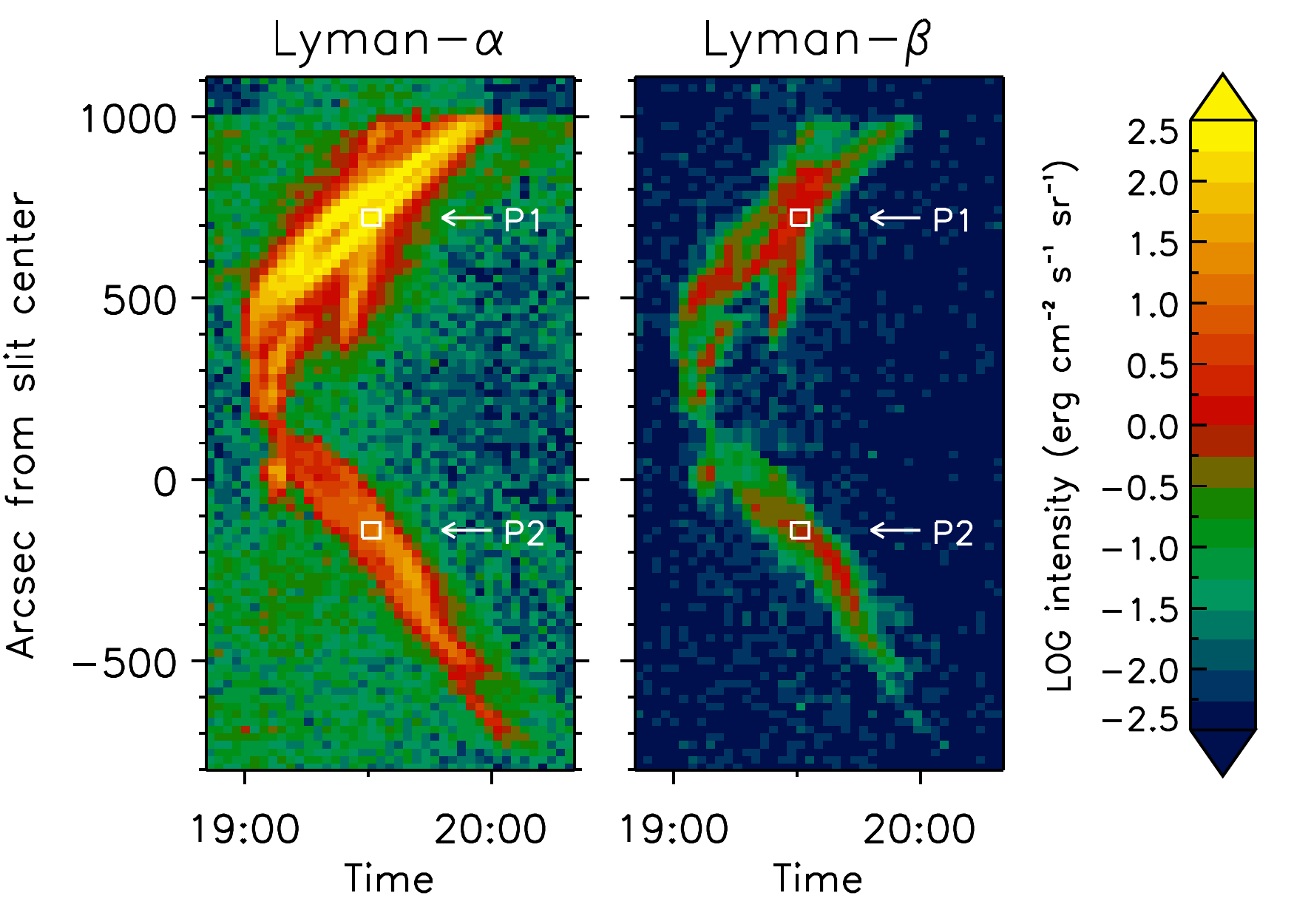}
\caption{\label{fig:int_maps}Intensity distribution of the narrow component of the prominence \lya\ and \lyb\ lines, plotted as a function of time and position along the UVCS slit. Note that $1\ \text{arcsec}\simeq800$~km.}
\end{figure}

UVCS recorded spectra of several UV coronal lines (see Table~1 of \paperone\ for a complete list) as the prominence crossed its field of view (FOV), a 40~arcmin long slit placed perpendicularly to the solar radius at a latitude of $40^\circ$NE and heliocentric distance of 2.3~\rsun\ (see Figure~\ref{fig:lasco}).
In particular, spectral profiles of the neutral hydrogen \mbox{Lyman-$\alpha$} ($\lambda_{12}=1215.67$~\AA) and \mbox{Lyman-$\beta$} ($\lambda_{13}=1025.72$~\AA) lines were acquired with spatial resolution of 21~arcsec ($\approx 15000$~km) and integration time of 120~s during all the CME event.
Both lines systematically appear to be the superposition of two components with different widths, the narrower one representing, in our interpretation, real prominence emission, while the broader one being more likely due to plasma in the PCTR or, alternatively, in the hot plasma shroud surrounding the prominence \citep[see][]{habbal2010}.

Separation of the two components was performed with a double Gaussian fit to the line profiles and its consistency was verified with a minimum chi-squared analysis (see \paperone).
A mean pre-event spectrum, obtained by averaging over several exposures preceding the onset of the eruption, was subtracted from the line profiles in order to remove from the observed emission the contribution coming from the quiet corona surrounding the prominence.
This correction may not be accurate because, as evidenced by the depleted (black) region around the prominence in Figure~\ref{fig:lasco}, much of the pre-event corona has been blown away by the CME during the eruption.
However, prominences are generally very bright in the Lyman lines \citep[e.g.,][]{ciaravella2003} and the quiet-corona contribution can be assumed to be very small; therefore possible uncertainties related to the background subtraction can be considered negligible in this case.

Figure~\ref{fig:int_maps} reports the total integrated intensities of the \lya\ and \lyb\ narrow components, as functions of time and position along the UVCS slit.
The prominence exhibits a complex structure in both lines, with ramifications and apparently superimposed emission features.
The intensity of the \lya\ line is up to 250 times higher than that of the \lyb.
The \lya\ to \lyb\ intensity ratio is consistent with the value somewhat lower than $\sim 500$ measured by \citet{ciaravella2003} in the prominence core of a CME observed with UVCS at an altitude of 2.3~\rsun.
Effective temperatures and Doppler shifts derived from the two line profiles, not shown here, outline the same scenario described in \paperone: the prominence appears to be almost uniform in temperature ($T_\text{eff}\approx10^5$~K), while striking differences in the LOS speeds characterize the two prominence legs, the southern one moving towards the observer at significantly higher velocity ($v_\text{LOS}\approx300$~km~s$^{-1}$) than the northern one, that, on the contrary, seems to remain anchored to the solar surface or slowly moving ($v_\text{LOS}\approx25$~km~s$^{-1}$).

\begin{table}
\caption{Observed quantities and modeling results.}\label{tab:results}
\centering
\begin{tabular}{lcc}
\hline\hline\noalign{\vskip 2pt}
                                                  & P1            & P2 \\
\hline\noalign{\vskip 2pt}
Heliocentric distance ($R_\odot$)                 & 2.46          & 2.35 \\
$N_e$ ($10^{17}$~cm$^{-2}$)                       & $1.48\pm0.03$ & $0.39\pm0.02$ \\
$I^{(\alpha)}$ (erg~cm$^{-2}$~s$^{-1}$~sr$^{-1}$) & $216\pm1$     & $12.5\pm0.8$ \\
$I^{(\beta)}$ (erg~cm$^{-2}$~s$^{-1}$~sr$^{-1}$)  & $0.9\pm0.3$   & $0.4\pm0.1$ \\
$T_\text{eff}$ ($10^4$~K)                         & $10.9\pm0.1$  & $8.7\pm0.8$ \\
$v_\text{LOS}$ (km~s$^{-1}$)                      & $24.6\pm0.3$  & $297\pm2$ \\
$v_\text{POS}$ (km~s$^{-1}$)                      & \multicolumn{2}{c}{$280\pm30$} \\
\hline\noalign{\vskip 2pt}
$p$ ($10^{-3}$~dyn~cm$^{-2}$)                     & 0.52          & 2.1 \\
$n_e$ ($10^{8}$~cm$^{-3}$)                        & 0.36          & 1.1 \\
$T$ ($10^4$~K)                                    & 4.96          & 6.67 \\
$T_\text{eff}$ ($10^4$~K)                         & 6.50          & 8.74 \\
$v$ (km~s$^{-1}$)                                 & 156           & 210 \\
$\tau^{(\alpha)}$                                 & 0.97          & 0.07 \\
$D_\text{eff}$ (km)                               & 40000         & 3600 \\
$f$ (\%)                                          & 71            & 7 \\
\hline
\end{tabular}
\end{table}

The observational quantities derived from visible-light and UV data were used in \paperone\ to constrain a non-LTE (i.e., departures from local thermodynamic equilibrium) radiative-transfer model of the prominence, which provided us with all the other thermodynamic plasma parameters, such as the gas pressure, kinetic temperature, microturbulent velocity, ionization degree, and line opacity.
The full non-LTE multilevel radiative-transfer problem was solved by means of a 1D numerical code (see \paperone\ for more details) for a set of prominence pixels, and, in particular, for the two sole points along the prominence where simultaneous and cospatial UVCS and LASCO data were available (points P1 and P2; see Figures~\ref{fig:lasco} and~\ref{fig:int_maps}).
The plasma parameters obtained for the two points are listed in the top part of Table~\ref{tab:results}, together with the statistical uncertainties that have been derived from the Gaussian-fit parameters.

For these two points, the code was iteratively run taking the measured quantities as initial input parameters and the gas pressure as a free parameter, until when the computed intensities of \lya\ and \lyb\ lines matched the observed values.
The additional information on the electron column density derived with LASCO was used to constrain the model, so that it was possible to derive even the effective prominence thickness, $D_\text{eff}$.
Calculations in \paperone\ were originally performed using the \lya\ and \lyb\ incident radiation profiles reported in \citet{gouttebroze1993}; they were obtained from solar disk measurements performed in 1976 (i.e., around solar activity minimum) with the OSO-8/LPSP instrument (see Sect. 5.2 of \paperone).
However, for the purposes of this work and in order to explore a different choice of the incident radiation profiles, the detailed modeling in points P1 and P2 has been repeated using the more recent \lya\ and \lyb\ line profiles measured on the solar disk with SOHO/SUMER in May 2000 and reported in \citet{lemaire2015} (see the details in the next section).
The final model parameters are listed in the second part of Table~\ref{tab:results}; they only slightly differ from the values reported in \paperone.

Beside the general results concerning the plasma thermodynamics that have been already discussed in \paperone\ and \papertwo, we remind here that the resulting effective thickness was combined with the observed thickness estimated from LASCO-C2 images, \mbox{$D=56000$~km}, to infer the prominence LOS filling factor in the two points.
This was done by implicitly assuming an irregularity factor $X=1$ (so that $f=D_\text{eff}/D$; see \paperone), because the 1D-slab prominence model assumes that all the plasma parameters are uniform along the LOS.
The filling factor obtained in this way turns out to be lower than one in both points.
This suggests that the prominence most likely undergoes fragmentation during its expansion; for instance, in point P2, where $f\ll1$, only a very small fraction of the LOS concurs to the Lyman emission observed by UVCS.
This supports the interpretation that in the southern prominence leg, which is moving and expanding at higher velocity, the emitting plasma may be more rarified and distributed on spatial scales smaller with respect to the length $D$ estimated from the visible light.

\section{Analysis methods}\label{analysis}

Aim of this work is to test a partially independent technique to infer the prominence LOS filling factor, based on UVCS observations of \lya\ and \lyb\ emission lines, and to check the consistency of the results against the model values presented in \paperone.
The general approach is to use the results of the detailed modeling of prominence points P1 and P2 (i.e., hydrogen kinetic and effective temperatures and outflow velocities) to (1) evaluate the expected intensities of the radiative and collisional components of one or both the hydrogen Lyman lines in the two points, according to the formalism described in Section~\ref{theory}, and (2) estimate the filling factor from the comparison with the measured intensities. 
All the other quantities are constrained by the observations or reasonably assumed.

\subsection{Incident radiation and Doppler-dimming calculations}\label{analysis:dimming}
\begin{figure}
\includegraphics[width=\columnwidth]{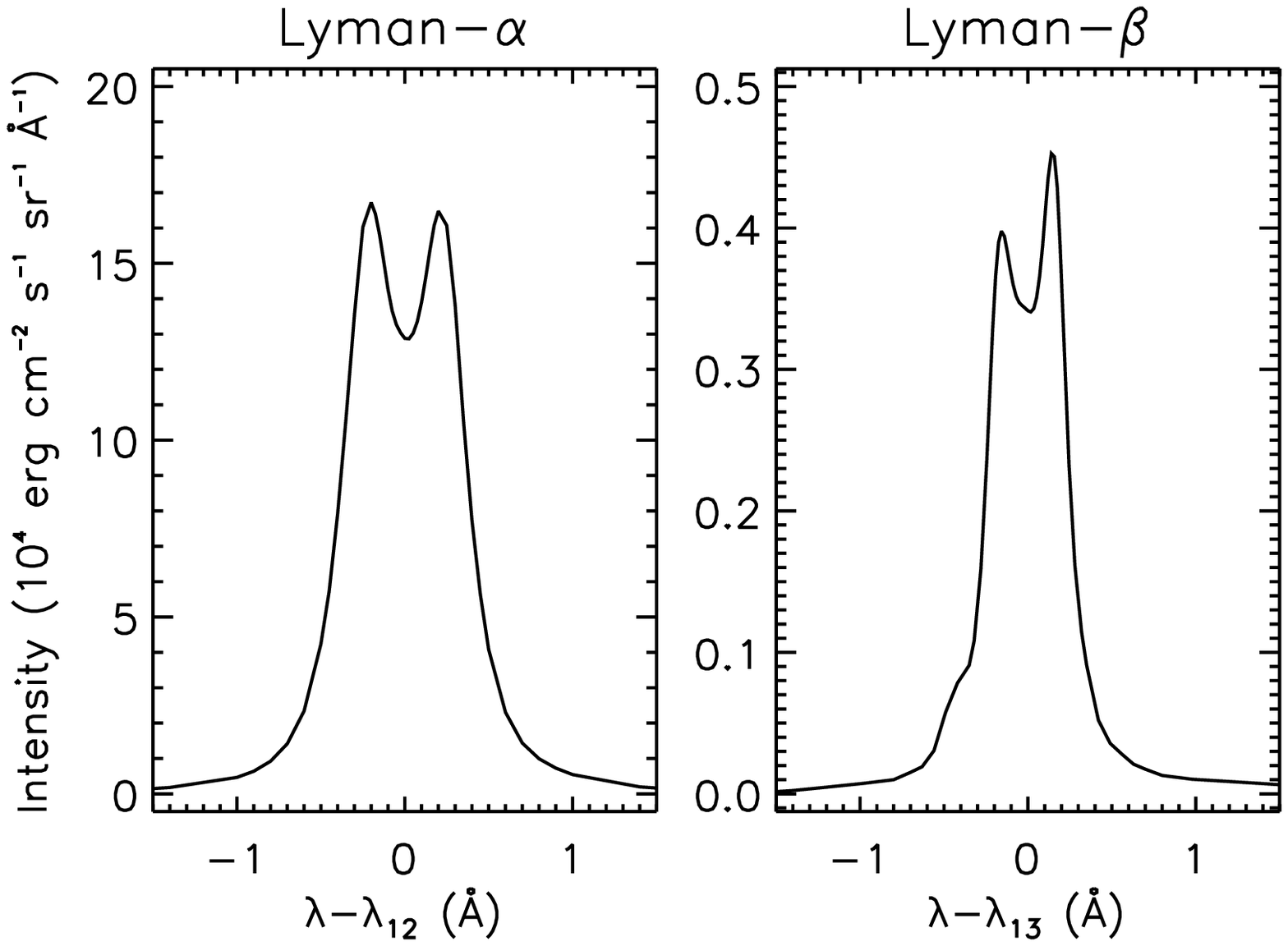}
\caption{\label{fig:profiles}Solar disk \lya\ and \lyb\ spectral profiles derived from SOHO/SUMER measurements \citep[see][]{lemaire2015}.}
\end{figure}
As anticipated in the previous section, the incident radiation profiles of both \lya\ and \lyb\ lines adopted for the present calculations were derived from the high-resolution spectral irradiance profiles measured on the solar disk with SOHO/SUMER on May 20, 2000 \citep[see][]{lemaire2015}; among the several measurements reported in that paper, spanning most part of the solar cycle 23, we selected the ones closest to the date of our event. 
Using the same approach described in \paperone, the \lya\ profile was renormalized to match the actual total integrated intensity at the time of the event, \mbox{$I^{(\alpha)}_\odot=1.29\times 10^5$~erg~cm$^{-2}$~s$^{-1}$~sr$^{-1}$}, derived from the \lya\ flux at the Earth measured with SOLSTICE; the \lyb\ profile was renormalized according to the results reported by \citet{lemaire2015} on the variations of the \lya\ to \lyb\ intensity ratio during the solar cycle, giving \mbox{$I^{(\beta)}_\odot\simeq 0.022\cdot I^{(\alpha)}_\odot$}.
The line profiles are shown in Figure~\ref{fig:profiles}.

\begin{figure}
\includegraphics[width=\columnwidth]{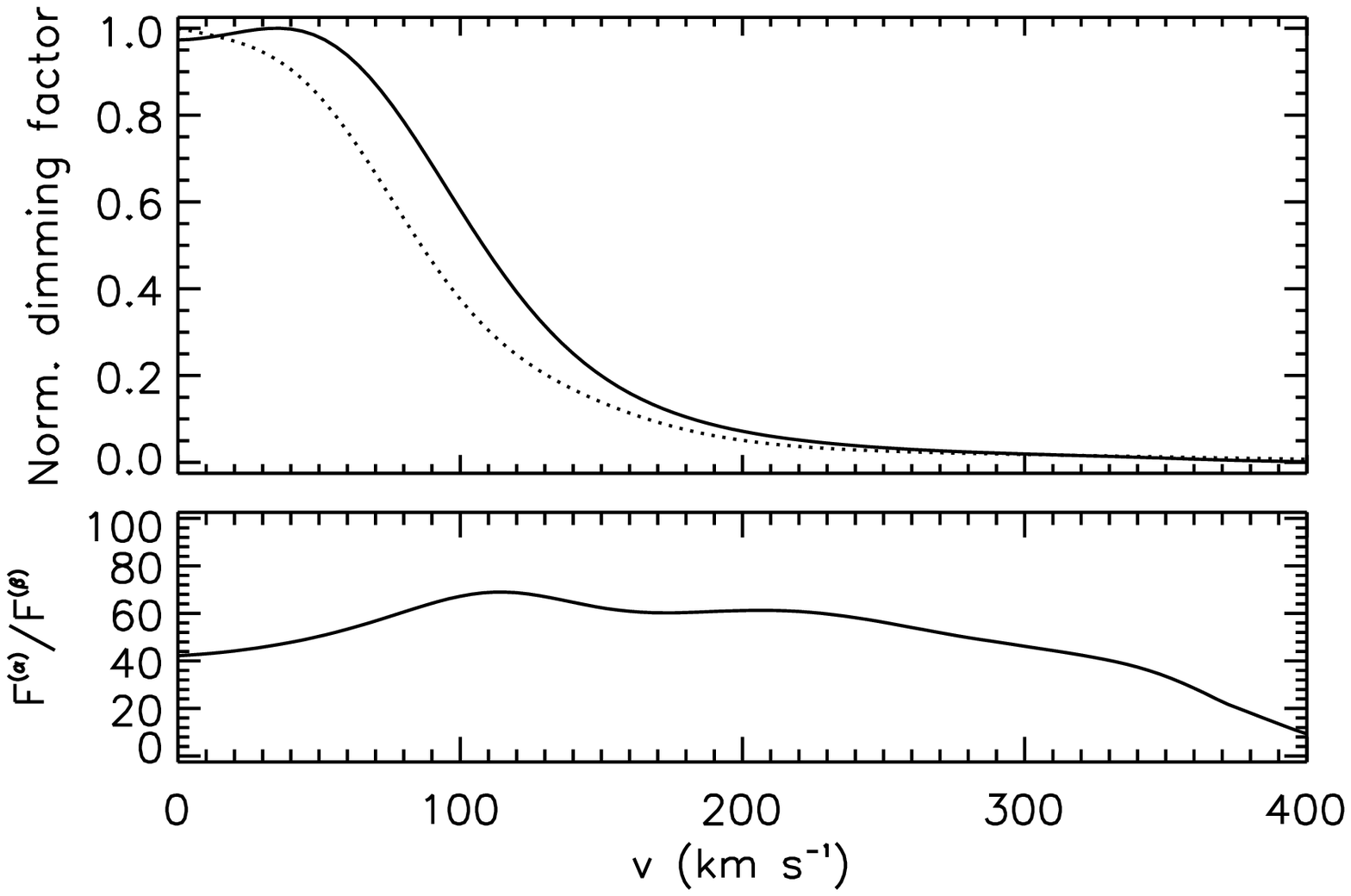}
\caption{\label{fig:dimming}Normalized Doppler-dimming factor of the hydrogen \lya\ (top panel, solid line) and \lyb\ (dotted line) lines, calculated for the representative temperature of $10^5$~K,} and ratio of two factors (bottom panel), as functions of the plasma outflow velocity.
\end{figure}

The Doppler-dimming factor is calculated assuming a Gaussian absorption profile with a FWHM corresponding to the effective temperatures in points P1 and P2 derived from the model and listed in Table~\ref{tab:results}.
The incident radiation profile has been red-shifted according to the outflow velocities ($v$) reported in the Table.
We remind that those velocity values are lower than the POS component of the prominence speed measured with LASCO, because they were required in order to get the best agreement between the observed and the synthesized \lya\ and \lyb\ intensities.
On the other hand, as described in \paperone, the analysis of UVCS \ion{O}{vi} data has put into evidence that the $v_\text{POS}$ estimated from LASCO images is most probably an overestimate of the real outflow velocity due to the prominence acceleration in the first phases of its evolution.

Figure~\ref{fig:dimming} shows, as an example, the dimming factors $F^{(\alpha)}$ and $F^{(\beta)}$ of the \lya\ and \lyb\ lines, respectively, computed as functions of the outflow velocity for the representative temperature of $10^5$~K, together with the ratio $F^{(\alpha)}/F^{(\beta)}$.
It is worth noting that the \lya\ dimming factor is not a monotonically decreasing function of the outflow velocity, at variance with the typical behavior obtained for coronal plasma \citep[see, for comparison,][]{kohl1982}.
This is caused by the relatively narrow absorption profile, in this case relevant to temperatures typical of prominences, making the Doppler dimming much more sensitive to the shape of the incident radiation profile.
In fact, the slight increase in the dimming factor occurring for velocities from zero to $\sim 40$~km~s$^{-1}$ is due to the transition from the central dip to the peak of the disk profile that actually leads to a ``pumping'' of the radiative component of the line, that is expected to be enhanced by the plasma flow in this velocity range.

\subsection{Collisional rates and ionization balance}
\begin{figure}
\includegraphics[width=\columnwidth]{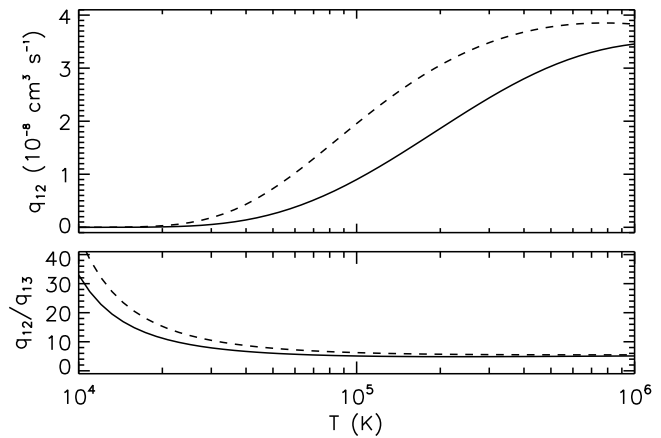}
\caption{\label{fig:qcol} Collisional coefficients for the \lya\ transition (top panel) and ratio of the \lya-to-\lyb\ coefficients (bottom panel), computed as functions of the temperature using the five-level hydrogen model atom with continuum described in \citet{gouttebroze1993} (solid lines) and the theoretical approximation (given by Equation~[\ref{eq:qcol}]; dashed lines).}
\end{figure}
The collisional excitation coefficients for \lya\ and \lyb\ transitions were numerically computed using the five-level hydrogen model atom with continuum described in \citet{gouttebroze1993}.
The resulting collisional rates for the \lya\ ($q_{12}$) are plotted in Figure~\ref{fig:qcol} vs. temperature.

It is interesting to compare these rates with an analytical approximation widely used for typical coronal lines, as found, for instance, in \citet{mewe1972}:
\begin{equation}\label{eq:qcol}
q_{ij}(T)\simeq2.73\times10^{-15}\,\frac{f_{ij}\cdot g}{E_{ij}\,T^{1/2}}\exp\left(-\frac{E_{ij}}{k_B T}\right),
\end{equation}
where $f_{ij}$ is the absorption oscillator strength for the transition, $g$ the average electron-impact Gaunt factor, $E_{ij}=hc/\lambda_{ij}$ the transition energy, and $q_{ij}$ is in c.g.s. units.
At the temperatures characteristic of the prominence plasma ($\sim10^5$~K), the collisional rates computed with this theoretical approximation are a factor of $\sim 2$ larger than the coefficients computed numerically with the five-level hydrogen atom model (see Figure~\ref{fig:qcol}).
Note that the collisional coefficient for the \lyb\ transition is in both cases about $\sim 20\%$ of that relevant to the \lya\ above $10^5$~K.

The hydrogen ionization fraction $R(T)$ is taken from the ionization equilibrium of \citet[][provided by the CHIANTI atomic database, version 7]{arnaud1985}.
Although this is a simplification, because plasma in expanding prominences can be out of ionization equilibrium, we checked that the ionization degree computed by the numerical code used in \paperone\ was in reasonable agreement with that provided by \citet{arnaud1985}.

\section{Results}\label{results}
According to Eq.~(\ref{eq:ff_calc}), in order to derive the prominence plasma filling factor it is necessary to separate the radiative and collisional components of one of the two Lyman lines.
To do that we follow two different methods, with different degrees of approximation.
In the first one, we use intensities of both \lya\ and \lyb\ lines to disentangle the radiative and collisional components of each line; in the second method, we evaluate the two components of the \lya\ intensity in points P1 and P2 as a function of the filling factor, using Equations~(\ref{eq:i_rad_approx}) and~(\ref{eq:i_col_approx}), and find the best value of $f$ from the comparison with the measured intensity.

\subsection{First method}\label{analysis:first_approach}
Given the total intensities of both \lya\ and \lyb\ lines, the collisional and radiative components can be easily separated according to \citet{fineschi1994} by using the ratios:
\begin{equation}\label{eq:r_col}
R_\text{col}\equiv\frac{I^{(\alpha)}_\text{col}}{I^{(\beta)}_\text{col}}=\frac{\lambda_{13}\,b_{12}\,q_{12}}{\lambda_{12}\,b_{13}\,q_{13}}\simeq0.96\cdot\frac{q_{12}}{q_{13}}
\end{equation}
and
\begin{equation}\label{eq:r_rad}
R_\text{rad}\equiv\frac{I^{(\alpha)}_\text{rad}}{I^{(\beta)}_\text{rad}}=\frac{B_{12}\,\lambda_{12}\,b_{12}\,F^{(\alpha)}}{B_{13}\,\lambda_{13}\,b_{13}\,F^{(\beta)}}\simeq8.4\cdot\frac{F^{(\alpha)}}{F^{(\beta)}},
\end{equation}
where all quantities are known or can be estimated as explained in the previous sections.
Using these relationships together with Eq.~(\ref{eq:i_tot}), Eq.~(\ref{eq:ff_calc}) can be rewritten as:
\begin{equation}\label{eq:ff_calc_2}
f=K(T)\cdot\frac{N_e^2}{D}\cdot\left[I_\text{tot}^{(\alpha)}-R_\text{rad}(v)\cdot\frac{I_\text{tot}^{(\alpha)}-R_\text{col}I_\text{tot}^{(\beta)}}{R_\text{rad}(v)-R_\text{col}}\right]^{-1}.
\end{equation}

The advantage of using the above expressions is that the separation of the radiative and collisional components of the two lines is based on relative ratios that are less sensitive to the overall assumptions (e.g., they are independent on the hydrogen ionization degree, the electron density, and the geometry of the resonant scattering); however, the reliance of the ratios on the collisional coefficients and the incident radiation profiles through the dimming factors is still a potential source of uncertainty.

\begin{table}
\caption{Resulting radiative and collisional components.}\label{tab:results_2}
\centering
\begin{tabular}{llcccc}
\hline\hline\noalign{\vskip 2pt}
              &                & \multicolumn{2}{c}{P1} & \multicolumn{2}{c}{P2} \\
              &                & \lya\ & \lyb\          & \lya\ & \lyb\ \\
\hline
\multirow{2}{*}{First method}  & $I_\text{rad}$ & 212.5 & 0.43           & 10.0  & 0.02 \\
              & $I_\text{col}$ & 3.5   & 0.47           & 2.5   & 0.38 \\
\hline
\multirow{2}{*}{Second method} & $I_\text{rad}$ & 183.2   & 0.55           & 5.6   & 0.01 \\
              & $I_\text{col}$ & 32.8    & 0.35           & 6.9   & 0.39 \\
\hline
\end{tabular}
\end{table}

According to the previous equations, the radiative component of the \lya\ line turns out to be $\sim 98\%$ of the total intensity in point P1 and~$\sim 80\%$ in point P2 (see Table~\ref{tab:results_2}).
Although the resonantly scattered component may be dimmed by more than 50\% at velocities above 100~km~s$^{-1}$ \citep[see Figure~\ref{fig:dimming} and, e.g., the discussion in][]{kohl1982}, its contribution is still overwhelming in both points.
The \lyb\ line, conversely, is more collisional, since the collisional component is $\sim 50\%$ of the observed intensity in P1 and up to $\sim 95\%$ in P2 (see Table~\ref{tab:results_2}).
This is in agreement with the general evidence that in high-density coronal structures, such as prominences or streamers, the \lya\ line is essentially radiatively formed while the \lyb\ line is more collisionally driven \citep[][]{labrosse2006,vial2016}.

The resulting filling factors are $f=(74\pm 21)\%$ in point P1 and $f=(5\pm 1)\%$ in point P2.
The uncertainties have been estimated from the measurement errors, in particular from the UVCS intensity errors.
We point out that actual uncertainties on the results may be larger, especially because of the large uncertainty affecting the column density measurement, that is, however, hardly quantifiable.
As we will also show in the following sections, small changes in the outflow velocity and kinetic temperature assumed for the calculations may dominate the uncertainty affecting the resulting filling factor, as well.

In both points we obtain a satisfactory agreement with the results from the NLTE model.
This is interesting and shows that even the simple method described here is able to give results similar to those provided by the more robust NLTE modeling when the necessary plasma parameters are well constrained from the observations.
However, it must be emphasized that our approach is based on relationships that are strictly valid for optically thin lines ($\tau< 1$), condition that is usually satisfied in the corona but not necessarily in erupting prominences, where densities are larger and temperatures lower. 
This nice agreement is then at least in part consequence of the mild optical depth characterizing the two selected points (see Table~\ref{tab:results}), compared to typical $\tau$ values derived at the center of the \lya\ and \lyb\ lines in quiescent prominences \citep[e.g.,][]{gouttebroze1993}.

The above results were obtained using the numerically-computed collisional rates; when the theoretical approximation of Eq.~(\ref{eq:qcol}) is used instead, an unrealistic filling factor \mbox{$f\gg1$} is obtained in point P1, while in point P2 we get $f=(18\pm 8)\%$, that is consistent with the model value only within~$2\sigma$.

\subsection{Second method}\label{analysis:second_approach}

\begin{figure}
\includegraphics[width=\columnwidth]{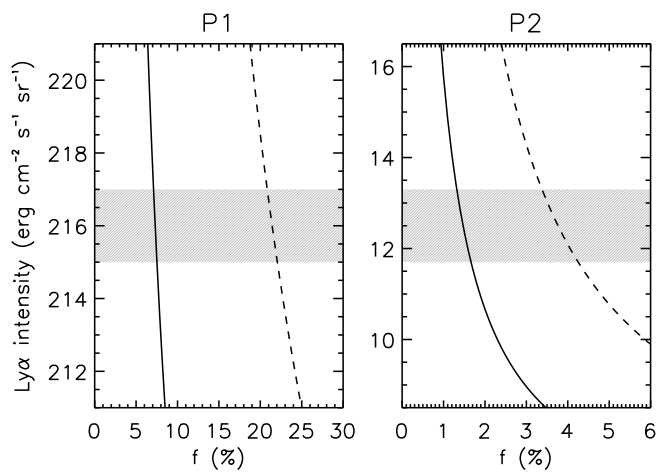}
\caption{\label{fig:radcol_lya} Total \lya\ intensity computed as a function of the LOS plasma filling factor using the method described in Section~\ref{analysis:second_approach} with the numerical (solid lines) and theoretical (using Eq.~[\ref{eq:qcol}]; dashed line) collisional coefficients for the computation of the collisional rates. The grey bands, plotted for reference purposes, mark the uncertainty range around the total integrated line intensities measured with UVCS.}
\end{figure}

We also tried to estimate the prominence filling factor using the intensity of the \lya\ line alone, in combination with the results obtained from LASCO-C2 data.
This attempt is motivated by the fact that we want to check the reliability of this method in the perspective of applying it to the future data delivered by the Metis coronagraph \citep[see][]{antonucci2012} on board the Solar Orbiter spacecraft \citep[see][]{muller2013} and other similar instruments such as the LST on board the ASO-S chineese mission \citep[see][]{li2015}.
In particular, Metis will provide for the first time simultaneous and cospatial \lya\ and visible-light images of the solar corona above $\sim 1.6$~\rsun\ in the whole instrument field of view, offering interesting opportunities for the study of erupting prominences.
We remark, however, that Metis will lack spectroscopic observations, therefore additional assumptions on the effective and kinetic temperatures, derived in this work from UVCS observations, will be required.
Nevertheless, it will provide high-resolution, high-cadence image sequences that will allow better measurement of the outflow speed, to within the uncertainties related to the direction relative to the POS, that can be estimated from considerations on the position of the eruption.

The radiative component of the \lya\ line can be directly estimated using Equation~(\ref{eq:i_rad_approx}) with the plasma parameters listed in Table~\ref{tab:results}; the collisional component can be evaluated instead as a function of the filling factor using Equation~(\ref{eq:i_col_approx}).
Adding the two gives total expected intensity vs. filling factor; the resulting curves for the two points are plotted in Figure~\ref{fig:radcol_lya}.

It is evident that for both points the filling factor derived from the comparison of the expected intensity with the measured one is lower than the prominence model value.
We get $f=(7.3\pm 0.2)\%$ in point P1 (against $71\%$) and $f=(1.5\pm 0.2)\%$ in point P2 (against 7\%).
The radiative component of the \lya\ line turns out to be $\sim 85\%$ of the total observed intensity in point P1 and~$\sim 45\%$ in point P2 (see Table~\ref{tab:results_2}), i.e., the predicted collisional component is significantly greater than that estimated with the \lya/\lyb\ ratio, because Eq.~(\ref{eq:i_rad_approx}) most probably underestimates the intensity of the radiative component.
This could explain why the filling factor is~$\sim 10$ times lower than the model value in point P1 and~$\sim 5$ times lower in point P2.
Note that the filling factor would increase by a factor of~$~\sim 3$ if the collisional rates were computed using Equation~(\ref{eq:qcol}); nevertheless, the results would be still lower than the model values.

\begin{figure}[t]
\includegraphics[width=\columnwidth]{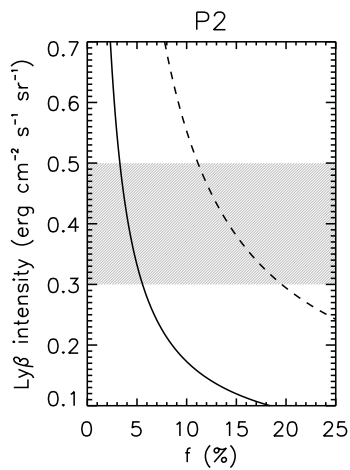}
\caption{\label{fig:radcol_lyb} As in Fig.~\ref{fig:radcol_lya}, for the \lyb\ intensity computed in point P2.}
\end{figure}

As a further test, we performed the same calculation using the \lyb\ line.
In point P1 we get \mbox{$f\approx1$}; conversely, in point P2 we obtain $f=(5\pm 1)\%$ using the collisional coefficients computed numerically (see Figure~\ref{fig:radcol_lyb}).
Given the large uncertainties affecting the intensity of this line (around $\sim30\%$, see Table~\ref{tab:results}), the rather acceptable agreement with the model achieved in point P2 is interesting.
To our opinion, two circumstances correspond to this result: (1) the collisional component of the \lyb\ intensity is dominant in point P2 ($\sim 95\%$ of the total intensity, in agreement with the value obtained with the first method; see Table~\ref{tab:results_2}), and (2) opacity effects in the \lyb\ line are very negligible in this point, because $\tau^{(\alpha)}\ll 1$ and, in general, $\tau^{(\beta)}<\tau^{(\alpha)}$ \citep[see, e.g.,][]{gouttebroze1993}.

\subsection{Exploration of the parameter space}\label{analysis:exploration}
In the previous sections we have used the measured Lyman line intensities and the visible-light brightness together with the plasma parameters ($v$ and $T$) derived from the NLTE prominence model to check the consistency of the methods employed to estimate the prominence plasma filling factor.
However, it is interesting to explore more extensively the parameter space, in order to collect information on the possible uncertainties of the computation and on the optimum ranges for the plasma parameter values.

\begin{figure}
\includegraphics[width=\columnwidth]{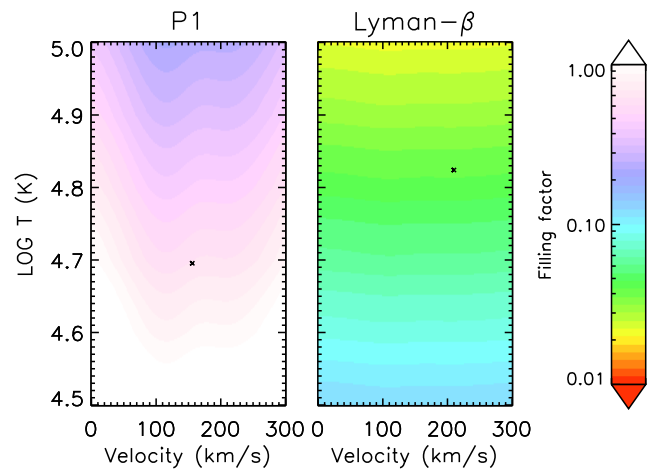}
\caption{2D maps of the filling factor obtained using Eq.~(\ref{eq:ff_calc_2}) as a function of outflow velocity and kinetic temperature, for point P1 (left panel) and P2 (right panel). The crosses mark the points in the $(v,T)$ plane corresponding to the NLTE-model results.\label{fig:ff_maps_1}}
\end{figure}

According to the method described in Section~\ref{analysis:first_approach}, Eq.~(\ref{eq:ff_calc_2}) can be used to compute the filling factor as a function of both outflow velocity and kinetic temperature, being the other plasma quantities (namely effective temperature and electron column density) fixed by the observations.
It is possible, therefore, to obtain 2D maps of the filling factor required to reproduce simultaneously the observed \lya\ and \lyb\ intensities in the $(v,T)$ space.

We limited our exploration to the velocity range between 0-300~km~s$^{-1}$, because, as found in \paperone, the prominence velocity derived from LASCO images ($v_\text{POS}\simeq 300$~km~s$^{-1}$) can be regarded as an upper limit to the real plasma outflow velocity.
As for the kinetic temperature, we considered the range between $10^{4.5}$-$10^5$~K, where the upper limit is constrained by the observed effective temperatures \mbox{($T_\text{eff}\lesssim10^5$~K)} while the lower one has been reasonably assumed according to the results of our analysis.
Figure~\ref{fig:ff_maps_1} shows the resulting filling-factor maps for the two points.

As it can be seen from the plots, the dependence of the filling factor on the plasma parameters turns out to be quite moderate.
In the explored domain, $f$ varies between $\sim 20$-100\% in point P1 and between $\sim 2$-12\% in point P2.
This ranges give a first-order indication on the possible values of the prominence plasma filling factor that are consistent with the NLTE model results.
Note that the dependence on the outflow velocity is weaker than that on the kinetic temperature, that turns out to be the most critical parameter. 
Interestingly, in point P2 the filling factor appears to be practically independent on the outflow velocity, and this is in agreement with the fact that in that point the line intensity is dominated by the collisional component. 

These results place important constraints on the use of the method described in Section~\ref{analysis:first_approach}; in fact, using Eq.~(\ref{eq:ff_calc_2}) it is possible to get an accurate estimate of the filling factor---or, at least, to  restrict the possible value range for this parameter---only if the plasma outflow velocity, and especially the kinetic temperature, are rather well constrained by the observations.
For instance, we estimated that an uncertainty $\Delta \log T=0.1$ corresponds to an average uncertainty of~$\sim 40$\% in the resulting filling factor.

\begin{figure}
\includegraphics[width=\columnwidth]{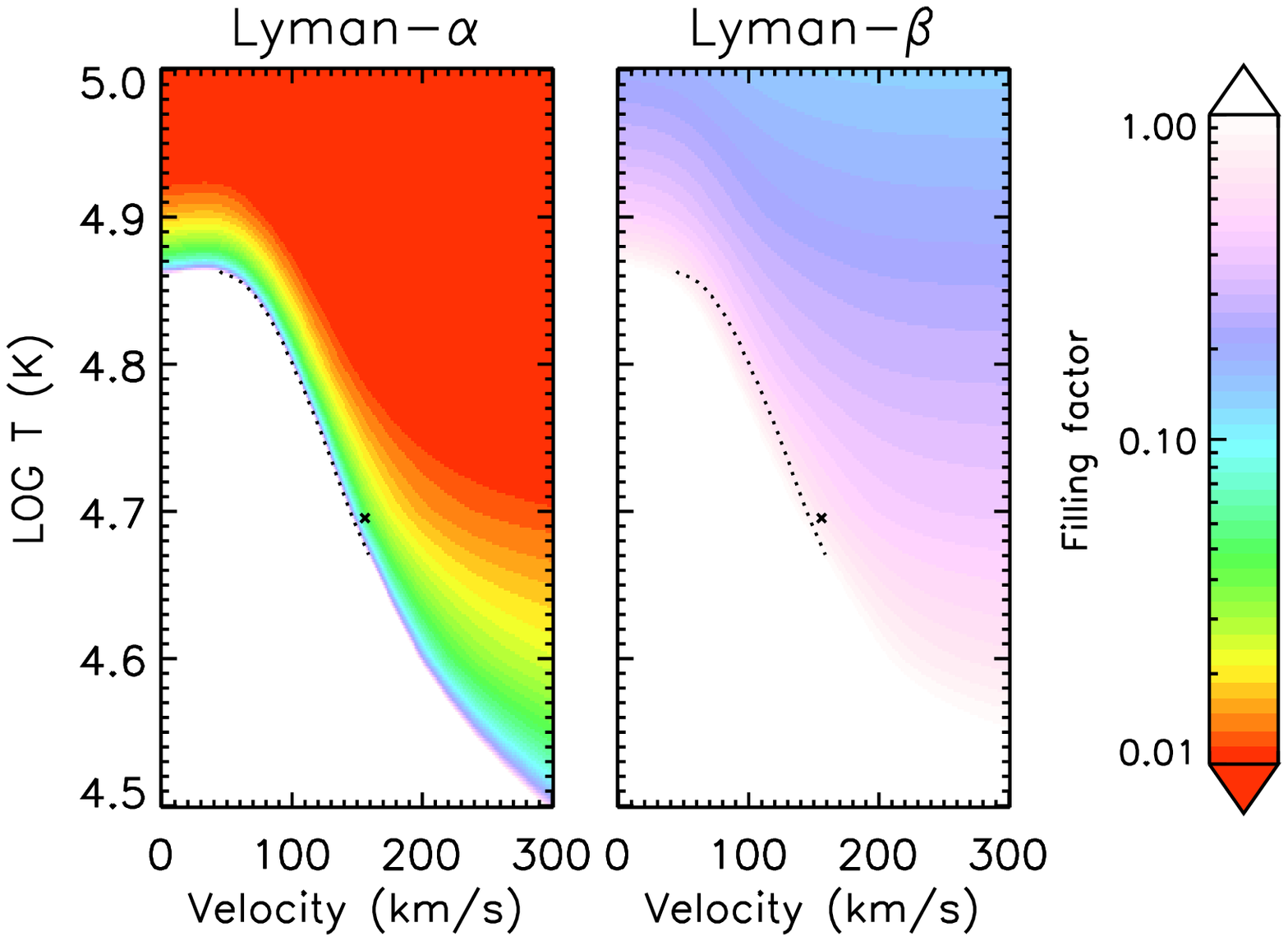}
\includegraphics[width=\columnwidth]{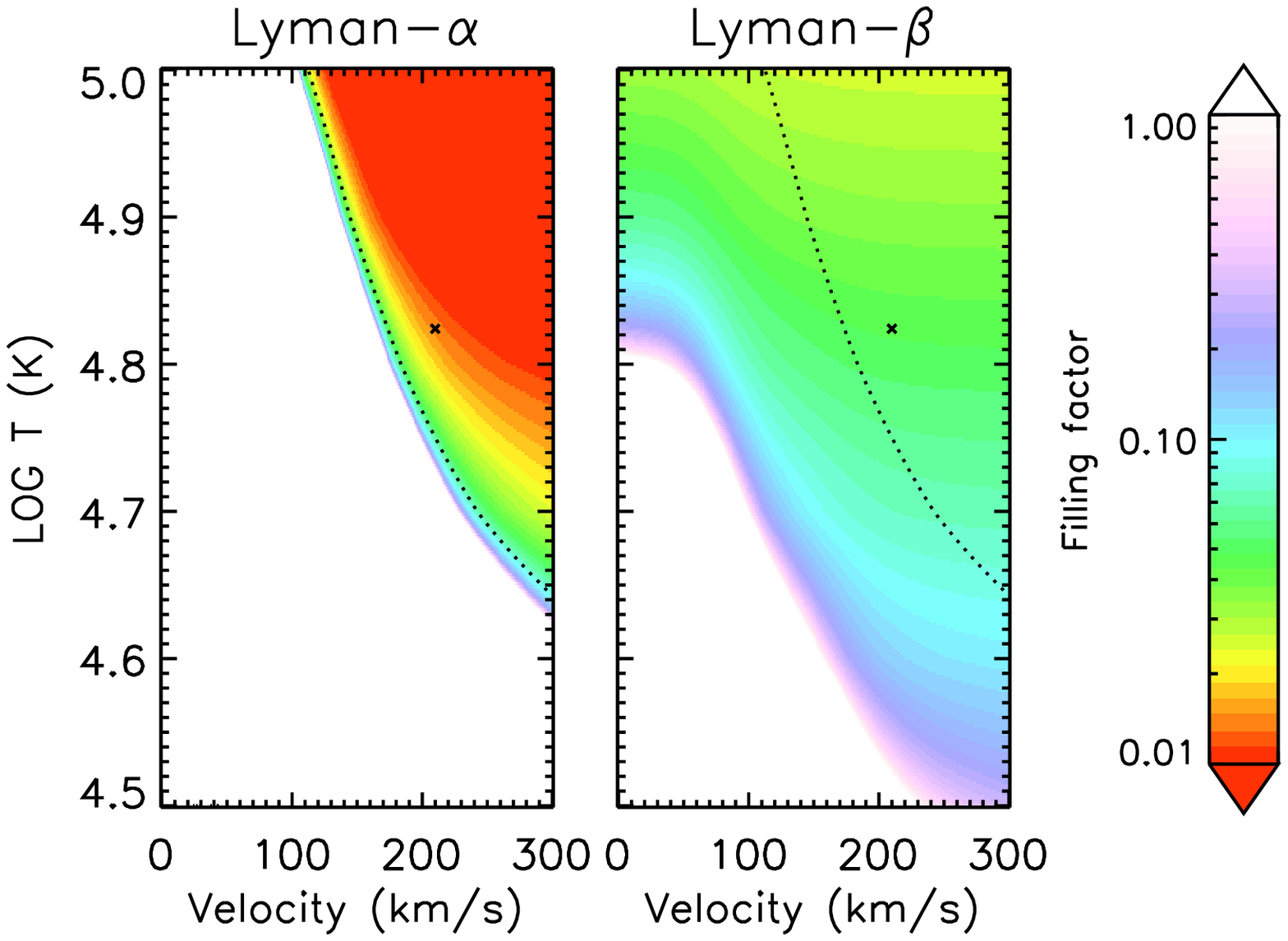}
\caption{Filling-factor maps obtained with combination of Eq.s~(\ref{eq:i_rad_approx}) and (\ref{eq:i_col_approx}), as function of outflow velocity and kinetic temperature. For each couple $(v,T)$, the maps give the corresponding $f$ values required to reproduce the intensities of the \lya\ (left panels) and \lyb\ (right panels) lines measured in point P1 (top panels) and P2 (bottom panels), respectively. The dotted line identifies the region of the parameter space (projected onto the $[v,T]$ plane) where intensities of both lines are matched simultaneously. As in Fig.~\ref{fig:ff_maps_1}, the crosses mark the points corresponding to the NLTE-model results.\label{fig:ff_maps_2}}
\end{figure}

Alternatively, using Equations~(\ref{eq:i_rad_approx}) and (\ref{eq:i_col_approx}), intensities of the \lya\ and \lyb\ lines can be calculated separately for each set of parameters $(v,T,f)$, i.e., the parameter space is three-dimensional in this case.
For both lines, the region of the parameter space where the computed intensity matches the observed one is therefore a 2D surface that can be represented as a color gradient map projected in the $(v,T)$ plane (see Figure~\ref{fig:ff_maps_2}).
The intersection of the two surfaces identifies the region where the intensity of both lines can be reproduced at once with a single set of parameters; this region is marked with the dotted line in the plots of Fig.~\ref{fig:ff_maps_2}.

The general trend is that the filling factor goes rapidly above 100\% for decreasing values of both outflow velocity and kinetic temperature; variations are steeper for the \lya\ line, whose intensity is more sensitive on the outflow velocity due to Doppler dimming.
In point P1, the region of the parameter space where simultaneous matching of the measured intensities is achieved is limited, so that it is possibile to reproduce the observed \lya\ and \lyb\ intensities only for velocities between 40-160~km~s$^{-1}$ and temperatures between $10^{4.67}$-$10^{4.87}$~K; the filling factor along the intersection region varies between 54-94\%.
It is worth noting that the point representative of the NLTE model (marked with a cross in the plots) lies outside this region, as expected on the basis of the results obtained in Section~\ref{analysis:second_approach}.
Although the approximative approach described there is not accurate at reproducing the model results when the two Lyman lines are considered separately, when they are combined together it is however possible to obtain a rough indication on the best ranges for the prominence plasma parameters that are in good agreement with the results of the detailed prominence modeling.

The situation is similar for point P2, for which, nevertheless, the parameter space region where the \lya\ and \lyb\ line intensities can be reproduced at once is not limited.
Therefore, the plasma parameters can be constrained in this case only by the observations, which suggest temperatures below $\sim 10^{4.9}$~K and velocities below~$\sim 300$~km~s$^{-1}$.
In this domain, the filling factor required to match both line intensities varies between 2-7\%.
This is again quite in agreement with the NLTE model results.

\section{Discussion and conclusions}\label{discussion}
Combination of cospatial and cotemporal visible-light and neutral hydrogen \lya\ and \lyb\ observations has been used in this work to obtain information on the LOS filling factor of the plasma embedded in an erupting prominence detected in the core of a CME.
The methods we presented are based on the comparison of the \lya\ and \lyb\ line intensities measured by UVCS with those evaluated using theoretical approximations for resonant scattering and collisional excitation, using the electron column density inferred from LASCO-C2 total-brightness data, and constraining all the other quantities from the observations or properly assuming them. 
In particular, we exploited the results of the detailed non-LTE prominence modeling described in \paperone\ to test the validity of these techniques.
The model provided us with the hydrogen kinetic temperature, microturbulence (and, therefore, the effective \lya\ and \lyb\ absorption line widths), and plasma flow velocity, in two prominence points where simultaneous and cospatial UVCS and LASCO data were available.
We derived the plasma filling factor with two slightly different methods, comparing the results with the values obtained from the model in order to check their reliability.

Our results show that the derived filling factors are satisfactorily consistent with the model values when the intensities of the \lya\ and \lyb\ lines are used together to disentangle the radiative and collisional components of each line, according to the technique described by \citet{fineschi1994}.
Since only the collisional component depends explicitly on the filling factor, the separation of the two components is crucial.
The use of two lines from the same atomic species---in our case, the hydrogen Lyman lines---minimizes the uncertainties due to the ionization balance, electron density, and element abundance, and reduces the effects relevant to the Doppler dimming and collisional excitation.

When the two lines are considered separately, the results are not consistent with the model, or there is only a marginal agreement in the prominence point where opacity effects are more negligible (i.e., point P2) due to the lower plasma optical depth.
On the one hand, this is a consequence of the fact that the approximate relationships used for the theoretical computation of the line intensities are very sensitive on the Doppler dimming coefficient, collisional rates, and hydrogen neutral fraction, thus the uncertainties on these terms may dominate depending on the predominant transition excitation regime (radiative/collisional). 
On the other hand, the formalism described in Section~\ref{theory} is strictly valid for UV emission in optically thin coronal lines.
As demonstrated in \paperone, a significant fraction of prominence points (between one third and one half) turn out to be optically thick in \lya, therefore one needs to be aware of the limitations of the results obtained when applying this technique to prominence observations.
The method described in Section~\ref{analysis:second_approach} may be then a better diagnostic for the outflow velocity \citep[see, e.g.,][]{dolei2018,bemporad2017} or the plasma temperature \citep[see][]{susino2016} than for the filling factor.

The most critical assumptions of our analysis are relevant to the incident radiation intensity and profile, the flow velocity, and the plasma temperature.
Intensity of the chromospheric \lya\ and \lyb\ radiation can be constrained by \lya\ irradiance measurements, while line profiles can be provided by solar disk spectral measurements.
The solar activity, either on global scales during the solar cycle \citep[see][]{tobiska1997} or locally in active regions, affects the intensities and the spectral profiles of both lines.
For instance, the \lya\ flux may increase by a factor of $\sim 1.5$ between solar minimum and maximum, and the \lyb\ by a factor slightly larger \citep[see][]{woods2000,lemaire2015}.
In addition, there is also evidence of the variability of the \lya/\lyb\ line-intensity ratio, depending on the predominant magnetic configuration (quiet Sun vs. active region) of the portion of the disk that illuminates the prominence \citep[see, e.g.,][and references therein]{tian2009}.
Therefore, the particular assumption made for the line profiles can affect the results to a large extent, the magnitude of these effects being related to the flow velocity through the Doppler-dimming term.

The radial component of the flow velocity can be roughly estimated from the the visible-light images, but this determination suffers of projection effects.
As we showed in \paperone, the POS component of the prominence velocity estimated from LASCO-C2 images is significantly larger than the radial component required to reproduce the observed intensities of the \lya\ and \lyb\ lines.
In our specific case, however, we find that the ratio of the Dimming factors $F^{(\alpha)}/F^{(\beta)}$ is slowly decreasing, within $\sim 30$\%, in the velocity range between 100-300~km~s$^{-1}$ (see Fig.~\ref{fig:dimming}), so that our final results are mildly dependent on the flow velocity when the method described in Section~\ref{analysis:first_approach} is used.

The knowledge of the hydrogen kinetic plasma temperature is fundamental to reliably evaluate the Lyman line intensities, and, in turn, the filling factor, especially in the temperature range typical of prominences, where collisional rates and ionization balance can vary by orders of magnitude. 
The kinetic temperature is related to the effective temperature through the (unknown) non-thermal motions. 
The sensitivity of our methods on the effective temperature, that constrains the width of the absorption profile, is quite weak---we checked that a variation of a factor of 2 on this parameter causes a variation no larger than $\sim 15$\% in the filling-factor values.
Conversely, the dependence on the kinetic temperature can be more strong, as evidenced by the exploration of the parameter space described in Section~\ref{analysis:exploration}, because this parameter affects both the collisional coefficients and the ionization fraction. 

Unlike the effective temperature, which can be constrained by the spectroscopic observations, the kinetic temperature must be inferred by means of assumptions on the non-thermal plasma motions that broaden the line profiles, such as the microturbulence. 
In \paperone, we assumed the microturbulent velocity to be a constant fraction of the sound speed (i.e., proportional to the square root of the plasma temperature), implying a precise relationship between kinetic and effective temperatures (see Appendix~A of \paperone).
Even if this hypothesis is plausible \citep[according to][]{parenti2007} and can be used for a first, approximative estimate of the microturbulent velocity, a detailed modeling of the hydrogen \lya\ and \ion{C}{iii} 977.02~\AA\ lines observed with UVCS (see \papertwo) shows that if these lines are used together to disentangle the kinetic temperature and the microturbulence from the line widths, the resulting microturbulent velocity in the prominence is not simply correlated to the kinetic temperature, but it is rather constant within the uncertainties (with values around 25~km~s$^{-1})$.
This makes it harder to rely the kinetic plasma temperature to the effective one.

These considerations underline that our results are significant as far as all the assumptions can be considered reasonable.
However, we also showed that combination of both the methods described in this work could be used not only to derive a first-order estimate of the prominence plasma filling factor, but also to provide an approximative, initial guess of the plasma parameters that could be used, in turn, to orient a more detailed modeling.
We point out that the use of simultaneous LASCO-C2 and UVCS data has restricted our analysis to only two prominence points, because of the low temporal cadence of C2 images and the limited spatial FOV of UVCS.
Therefore, additional investigation by considering, for instance, a number of eruptive prominence events observed in the visible light and UV would be useful to further test the techniques presented in this work.

\begin{acknowledgements}
Support from the Agenzia Spaziale Italiana through contract ASI/INAF No. I/013/12/0-1 is kindly acknowledged by RS and AB.
SJ and PH acknowledge the support from the grant of the Czech Funding Agency No. 16-18495S.
SJ also acknowledges financial support from the Slovenian Research Agency No. P1-0188.
We also thank the Referee for his valuable suggestions which helped to greatly improve the paper.

\end{acknowledgements}

\bibliography{bibliography}
\bibliographystyle{aa}
\end{document}